%% file: main.tex
\documentclass[sigconf]{acmart}
\AtBeginDocument{%
  }

\copyrightyear{2025}
\acmYear{2025}
\setcopyright{acmlicensed}
\acmConference[MM '25] {Proceedings of the 33rd ACM International Conference on Multimedia}{October 27--31, 2025}{Dublin, Ireland.}
\acmBooktitle{Proceedings of the 33rd ACM International Conference on Multimedia (MM '25), October 27--31, 2025, Dublin, Ireland}
\acmISBN{979-8-4007-2035-2/2025/10}
\acmDOI{10.1145/3746027.3758218}

\usepackage{multirow}
\usepackage{pifont}
\usepackage{xcolor}     
\usepackage{colortbl}
\usepackage{stfloats}

\begin{document}

\title{FineBadminton: A Multi-Level Dataset for Fine-Grained Badminton Video Understanding}

\author{Xusheng He}
\email{xsxshxs22@gmail.com}
\affiliation{%
  \institution{Harbin Institute of Technology, Shenzhen}
  \country{}
}

\author{Wei Liu}
\email{liuwei030224@gmail.com}
\affiliation{%
  \institution{Harbin Institute of Technology, Shenzhen}
  \country{}
}

\author{Shanshan Ma}
\email{mass1@cesi.cn}
\authornotemark[1]
\affiliation{%
  \institution{China Electronincs Standardization Institute}
  \country{}
}

\author{Qian Liu}
\email{leonaliuqian@gmail.com}
\affiliation{%
  \institution{Shandong University}
  \country{}
}

\author{Chenghao Ma}
\email{mach@cesi.cn}
\affiliation{%
  \institution{China Electronincs Standardization Institute}
  \country{}
}

\author{Jianlong Wu}
\email{wujianlong@hit.edu.cn}
\authornote{Corresponding authors.} 
\affiliation{%
  \institution{Harbin Institute of Technology, Shenzhen}
  \country{}
}

\renewcommand{\shortauthors}{Xusheng He et al.}

\begin{abstract}
Fine-grained analysis of complex and high-speed sports like badminton presents a significant challenge for Multimodal Large Language Models (MLLMs), despite their notable advancements in general video understanding. This difficulty arises primarily from the scarcity of datasets with sufficiently rich and domain-specific annotations.
To bridge this gap, we introduce FineBadminton, a novel and large-scale dataset featuring a unique multi-level semantic annotation hierarchy (Foundational Actions, Tactical Semantics, and Decision Evaluation) for comprehensive badminton understanding. 
The construction of FineBadminton is powered by an innovative annotation pipeline that synergistically combines MLLM-generated proposals with human refinement. 
We also present FBBench, a challenging benchmark derived from FineBadminton, to rigorously evaluate MLLMs on nuanced spatio-temporal reasoning and tactical comprehension. 
Together, FineBadminton and FBBench provide a crucial ecosystem to catalyze research in fine-grained video understanding and advance the development of MLLMs in sports intelligence.
Furthermore, we propose an optimized baseline approach incorporating Hit-Centric Keyframe Selection to focus on pivotal moments and Coordinate-Guided Condensation to distill salient visual information. 
The results on FBBench reveal that while current MLLMs still face significant challenges in deep sports video analysis, our proposed strategies nonetheless achieve substantial performance gains.
The project homepage is available at \url{https://finebadminton.github.io/FineBadminton/}.
\end{abstract}

\ccsdesc[500]{Computing methodologies~Computer vision}
\ccsdesc[500]{Information systems~Multimedia streaming}


\keywords{Sports Understanding, Multi-modal LLMs, Badminton Video Dataset}



\maketitle

\section{Introduction}
Multimodal Large Language Models (MLLMs) have demonstrated remarkable progress in general video understanding tasks~\cite{liang2024survey,hu2025video,li2024mvbench,fu2024video, wang-etal-2025-adaretake}. However, their capabilities are significantly challenged in scenarios demanding fine-grained analysis~\cite{ li2023mask, tang2024vidcompositionmllmsanalyzecompositions,he2025analyzing, li2023fine, ding2025rablip}, where subtle spatio-temporal details, complex event interactions, and domain-specific knowledge are paramount. This limitation is particularly pronounced in intricate sports like badminton, which is characterized by high-speed rallies, a diverse set of nuanced strokes executed with split-second timing, and rapid tactical decision-making~\cite{doi:10.1609/aaai.v38i8.28692,liu2025f3setanalyzingfastfrequent,chen2023tracknetv3}. These complexities push the boundaries of current video analysis far beyond coarse action recognition, requiring deeper and hierarchical gameplay comprehension~\cite{rao2024unisoccer,rao2024matchtimeautomaticsoccergame,wang2025haic}.

A critical bottleneck hindering advancement in this area is the scarcity of richly annotated datasets specifically designed for fine-grained badminton analysis. While general video datasets and some sports-specific collections exist, they often lack the necessary granularity or domain relevance. Although existing badminton resources~\cite{shuttle22,wang2023shuttleset,li2024videobadminton} are valuable for foundational tasks like stroke classification or shuttlecock tracking, they typically focus on more basic aspects and do not capture the deeper tactical intentions, precise execution nuances, or evaluative dimensions essential for comprehensive understanding. For instance, they may identify a "net shot" but fail to distinguish a strategically deceptive cross-court spinning net shot from a standard one, or to assess its impact on the rally. Manually annotating the dense, complex, and multi-layered events inherent in badminton at the necessary scale and detail is prohibitively expensive and prone to inconsistencies.

\begin{figure}[t]
    \centering
    \includegraphics[width=\linewidth]{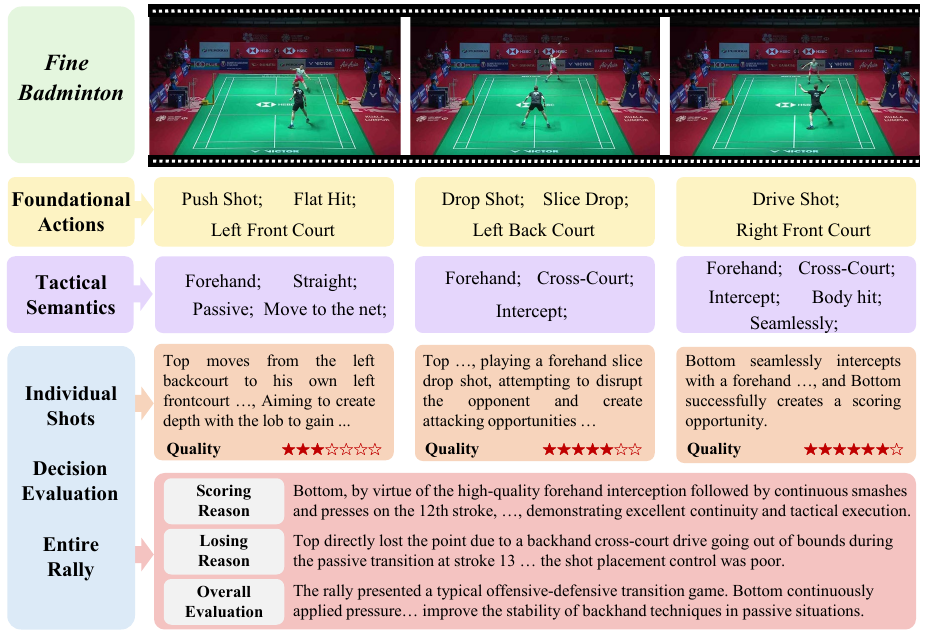}
    \vspace{-2.5em}
    \caption{An annotation example from FineBadminton illustrating its multi-level hierarchy. The figure showcases annotations for distinct shot moments within a single rally, detailing (1) Foundational Actions, (2) Tactical Semantics, and (3) Decision Evaluation.}
    \vspace{-1.5em}
\label{fig:dataset_overview}
\end{figure}

To bridge this critical gap, we introduce \textbf{FineBadminton}, a novel and large-scale dataset carefully designed for fine-grained badminton video understanding. FineBadminton features a unique multi-level annotation hierarchy providing unprecedented depth: (1) \textbf{Foundational Actions}, offering detailed classifications of stroke types and their execution subtleties; (2) \textbf{Tactical Semantics}, which describe ball trajectory dynamics and infer player strategic intentions, like deceptive plays or attacking formations; and (3) \textbf{Decision Evaluation}, providing expert-informed assessments of individual shot quality and entire rally narratives, including point-winning and error-inducing sequences. This hierarchical structure, exemplified in Figure~\ref{fig:dataset_overview}, facilitates a nuanced analysis by systematically detailing foundational actions, interpreting their tactical significance, and evaluating their contribution to the rally's outcome.

Recognizing the impracticality of purely manual annotation for such intricate detail, FineBadminton is developed using an innovative MLLM-driven automated annotation pipeline. This pipeline synergistically integrates foundational video models for structural parsing (e.g., player and ball tracking, hit event detection) with the advanced reasoning capabilities of MLLMs and LLMs to propose detailed and multi-level annotations. These proposals are subsequently refined by human experts, ensuring high-quality annotations with professional badminton insight. This human-in-the-loop approach significantly accelerates the creation of complex and hierarchical annotations while maintaining domain accuracy.

To validate FineBadminton's utility and promote further research, we establish \textbf{FBBench}, a comprehensive benchmark  designed to evaluate MLLM capabilities in nuanced and tactical sports video understanding. FBBench features a hierarchical task structure including Count, Action, Position and Cognition categories, assessing a model's capabilities in complex temporal reasoning, fine-grained action understanding, spatial awareness, and strategic inference within dynamic rally contexts. Furthermore, we propose an optimized baseline approach, underpinned by two novel badminton-specific strategies: Hit-Centric Keyframe Selection and Coordinate-Guided Information Condensation. This approach is designed to effectively adapt MLLMs to the demanding fine-grained tasks benchmarked by FBBench. We validate the effectiveness of these strategies through extensive experiments on FBBench, showing performance improvements across leading open-source and proprietary commercial MLLMs~\cite{gpt4.1,team2023gemini,doubao,zhang2025videollama,bai2025qwen2}.

Our contributions can be summarized as follows: 

(1) We introduce FineBadminton, a large-scale dataset of 3,215 rally clips (33,325 strokes), featuring an unparalleled multi-level semantic annotation hierarchy for in-depth badminton analysis beyond simple stroke recognition.

(2) We develop an innovative MLLM-driven Annotation Pipeline enabling scalable and consistent creation of fine-grained, hierarchical annotations with expert refinement.

(3) Building upon FineBadminton, we present FBBench benchmark, designed with structured tasks for rigorous evaluation of nuanced sports video understanding.

(4) We propose an optimized approach with novel strategies to effectively adapt MLLMs for fine-grained analysis on FBBench.

\section{Related Work}
\subsection{Video Understanding Datasets}
Video understanding datasets have rapidly evolved from early benchmarks focused on video classification~\cite{kay2017kinetics, tang2019coin} and action recognition~\cite{goyal2017something, caba2015activitynet}, to encompassing more complex tasks such as video question answering~\cite{lei2018tvqa, xu2017video}. Recent advancements have further driven the field towards finer-grained understanding, with datasets targeting sophisticated temporal reasoning~\cite{mangalam2023egoschema, lei2021detecting, li2024mvbench, wang2025fly} and the analysis of long-form videos~\cite{chen2024sharegpt4video, Farré2024FineVideo, yang2024vript}. However, while these general-purpose datasets significantly advance video intelligence, their application to specialized domains like sports is often constrained by limited domain-specific content and coarse-grained annotations, which lack the nuanced details crucial for in-depth analysis. These limitations highlight the need for fine-grained datasets tailored to complex and dynamic domains like sports.

\vspace{-0.5em}
\subsection{Sports Video Understanding Datasets}
Sports video understanding has emerged as a rapidly advancing research area, driven by its significant practical applications. Existing datasets in this domain can be broadly categorized into multi-sport collections~\cite{li2024sports, xia2025sportu,wu2024sportshhi} and sport-specific resources~\cite{rao2024unisoccer, held2024x, xu2024finesports,zhang2023logo,doi:10.1609/aaai.v38i8.28692,rao2024matchtimeautomaticsoccergame,liu2025f3setanalyzingfastfrequent}. Representative multi-sport datasets include Sports-QA~\cite{li2024sports} which encompasses eight sports using videos from action recognition datasets MultiSports~\cite{li2021multisports} and FineGym~\cite{shao2020finegym}. In contrast, sport-specific datasets offer more specific and fine-grained annotations, such as SoccerReplay-1988~\cite{rao2024unisoccer} and SoccerNet-XFoul~\cite{held2024x} for soccer analysis, and FineSports~\cite{xu2024finesports} targeting basketball activities.
In this work, we focus on badminton. Existing badminton datasets~\cite{shuttle22, wang2023shuttleset, li2024videobadminton} provide foundational stroke-level annotations, yet they lack the granularity, tactical semantics, and evaluative depth necessary for comprehensive understanding. 

To address these limitations, we introduce \textbf{FineBadminton}, featuring a multi-level annotation framework for badminton analysis, alongside \textbf{FBBench}, a comprehensive benchmark.

\vspace{-0.5em}
\section{FineBadminton: A Fine-Grained Badminton Video Dataset with Hierarchical Semantics}
To facilitate in-depth, fine-grained badminton gameplay analysis, we introduce \textbf{FineBadminton}, a novel large-scale dataset with comprehensive multi-level semantic annotations. This section outlines its construction methodology (Section~\ref{subsec:dataset_construction}), MLLM-driven annotation pipeline (Section~\ref{subsec:Pipeline}), and statistical analysis (Section~\ref{subsec:Statistics}).

\subsection{Dataset Construction}\label{subsec:dataset_construction}
FineBadminton is meticulously constructed through several key stages, beginning with data sourcing and culminating in comprehensive quality control, to ensure a reliable and high-quality dataset.

\textbf{Data Preparation.}  FineBadminton is curated from publicly accessible professional badminton match videos, officially released by the Badminton World Federation (BWF) on YouTube, covering events like the Olympics, World Championships, and various Open series events. The predominant camera perspective is an overhead view from behind the court, supplemented by a smaller subset of side-angle clips. All video content features  a resolution of 720p or higher, with a consistent frame rate of 25 FPS.

\textbf{Annotation Scheme.} To capture badminton's intricate semantic complexity, we design a multi-level hierarchical annotation scheme. This structure facilitates a progressive and comprehensive understanding of sports semantics across three abstraction levels: Foundational Actions, Tactical Semantics, and Decision Evaluation. All terminology and classifications within these levels are developed in collaboration with professional badminton players to ensure domain accuracy and relevance.

The \textbf{Foundational Actions} level focuses on the granular classification of individual badminton strokes, identifying the specific type of shot. Building upon the ShuttleSet22 framework and expert input, we deconstruct single strokes into the following 11 primary categories. These primary categories are further refined into 20 sub-categories based on the nuances of hand and wrist movements during execution. For instance, a `smash' can be sub-classified as a `jump smash' or a `slice smash'. Notably, a single stroke can belong to multiple sub-categories, enhancing descriptive granularity.

The \textbf{Tactical Semantics} level incorporates ball movement dynamics and inferred player intent, connecting physical actions to tactical awareness. The lexicon for this level, derived from common parlance in expert commentary and validated by professional players, includes distinctions based on trajectory (e.g., `straight line,' `cross-court'), position relative to the opponent (e.g., `passing shot,' `body shot,' `overhead shot'), and outcome relative to the court (e.g.,`out of bounds,' `hitting into the net'). Player intent reflects subjective elements such as `deception,' or situational assessments like `defensive play' and `passive/transitional shot.'

The \textbf{Decision Evaluation} level provides a higher-level and temporally extended assessment, evaluating individual shot quality and the overall rally's narrative. This level incorporates quality scores and textual descriptions for each shot. It broadens the perspective to an entire rally, by analyzing its characteristics, identifying key shots, and detailing reasons for points won or lost.  It also captures difficult-to-categorize in-game events (e.g., distinctive postures, critical movements) to ensure a holistic semantic representation.

\textbf{Annotation Process.} Given the complexity and hierarchical nature of our requirements, we develop custom GUI-based annotation software using Python. The annotation proceeds in two principal phases:(1) An initial phase relying entirely on manual annotation by trained human annotators.(2) A second phase where human annotators review and refine outputs generated by our automated annotation pipeline (detailed in Section~\ref{subsec:Pipeline}). This approach allows us to scale the annotation effort while maintaining high quality.

\textbf{Quality Control.} A stringent quality control protocol ensures the accuracy, consistency, and reliability of FineBadminton. The annotation team, comprising four experienced badminton enthusiasts, undergoes comprehensive training and completes pilot exercises before starting the main task. In the initial manual phase, each data sample is independently annotated by two individuals; a senior annotator then reviews these annotations, resolves discrepancies, and consolidates them into a unified and high-quality label. For the subsequent pipeline-assisted phase, outputs from the automated system are thoroughly reviewed and refined by two human annotators to ensure compliance with our established quality standards.

\subsection{Automated Annotation Pipeline}\label{subsec:Pipeline}

\begin{figure}[t]
    \centering
    \includegraphics[width=\linewidth]{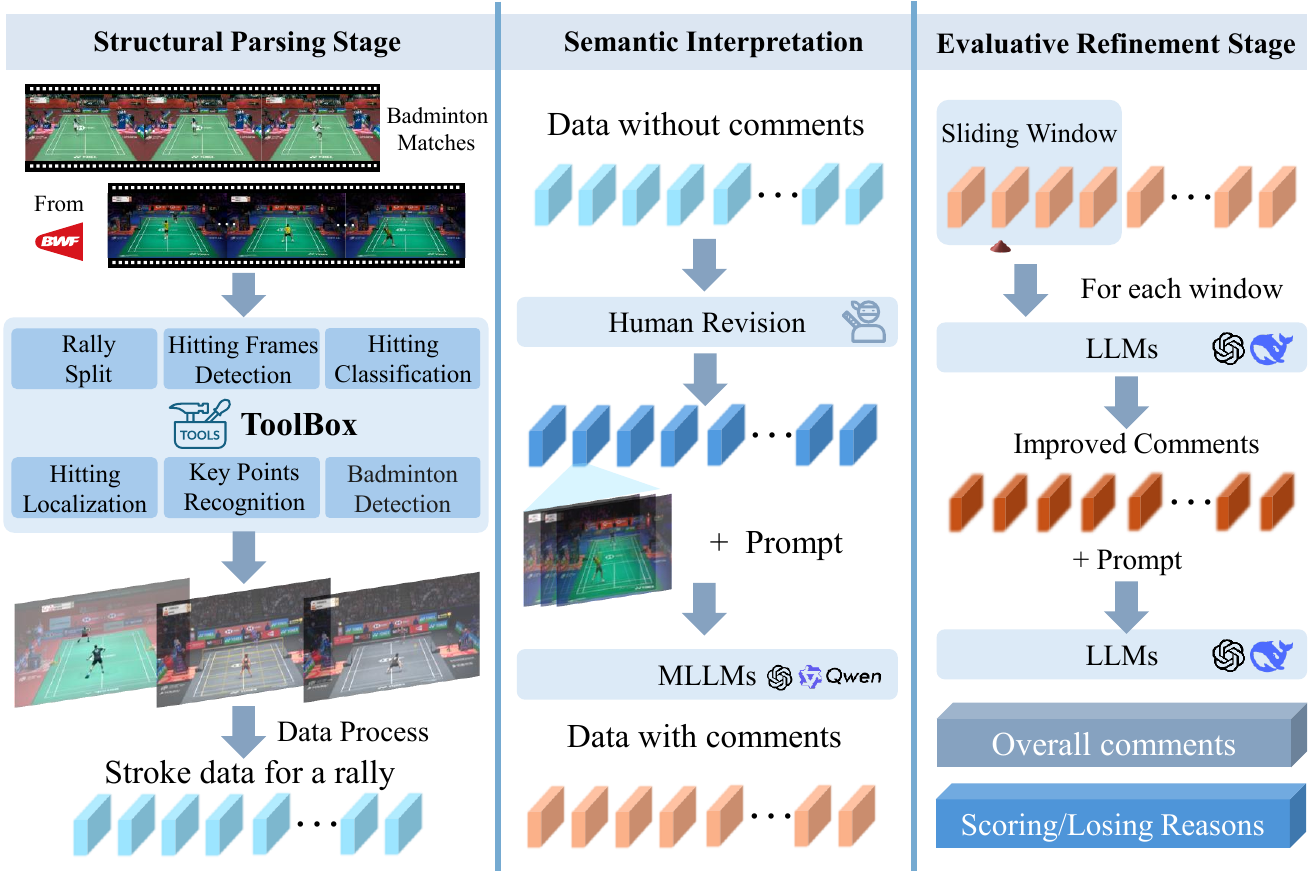}
    \vspace{-2em}
    \caption{Architecture of our automated annotation pipeline. The three-stage process first uses Structural Parsing to extract structured data from video, then Semantic Interpretation to generate descriptions for each stroke, and finally Evaluative Refinement to produce a cohesive rally evaluation.}
    \vspace{-2em}
\label{fig:pipeline_overview}
\end{figure}

To efficiently construct the FineBadminton dataset with rich and hierarchical annotations, we develop a multi-stage automated annotation pipeline, illustrated in Figure~\ref{fig:pipeline_overview}. The pipeline processes raw badminton match videos to extract structured information, interpret it semantically, and refine these interpretations into nuanced evaluations. The pipeline comprises three core stages: Structural Parsing, Semantic Interpretation, and Evaluative Refinement.

The initial \textbf{Structural Parsing Stage} deconstructs raw match footage into fundamental spatio-temporal events and attributes. This stage employs a "toolbox" of specialized models to extract key information crucial for badminton analysis. For ball trajectory detection, we employ TrackNetV3\cite{chen2023tracknetv3} due to its robustness against occlusion. Court boundary and player keypoint localization are achieved using FastRCNN\cite{girshick2015fast}. For identifying hit events (frames where a stroke occurs) and enabling deeper action understanding, we utilize VideoMAE\cite{DBLP:conf/nips/TongS0022,DBLP:conf/cvpr/WangHZTHWWQ23}, fine-tuned on badminton videos, as the visual feature extractor. These features, extracted from frames around the detected hits, are then combined with corresponding ball and player coordinate data and fed into a spatio-temporal feature learning and fusion module. This module enables task-specific prediction heads to classify stroke types and localize shuttlecock landing spots. The resulting structured data $\mathbb{D}$, containing elements like ball trajectories, player/court localizations, hit instances, stroke types, and landing spots, then proceeds to manual verification, as detailed in Section~\ref{subsec:dataset_construction}.

The second stage, \textbf{Semantic Interpretation}, translates the extracted structured data into rich textual narratives. This phase harnesses the advanced cross-modal reasoning capabilities of MLLMs. For each j-th identified stroke in a rally, its visual context $I_j$(three frames sampled around the hit instance) and associated structured data $D_j$ are fed into the MLLM. The model then generates a fine-grained textual description $T_j$ for that stroke,  characterizing and evaluating its specific qualities.

Finally, the \textbf{Evaluative Refinement Stage} produces contextually aware assessments of rally dynamics. While the previous stage provides detailed descriptions $T_j$ for each stroke, these outputs may lack rally context. To address this, a refining LLM processes each $T_j$. This refinement considers a contextual window composed of a total of w stroke descriptions, typically centered around the current stroke $T_j$, producing an improved $T'_j$. Subsequently, for a rally $R$ comprising a sequence of $N$ refined stroke descriptions $(T'_1, T'_2, \dots, T'_N)$, a comprehensive overall rally evaluation and reasons for scoring or losing is generated by a summarizing LLM. This final step ensures that the interplay between shots and the evolving tactical situation is more accurately captured in the narrative.

\begin{figure}[t]
    \centering
    \includegraphics[width=0.8\linewidth]{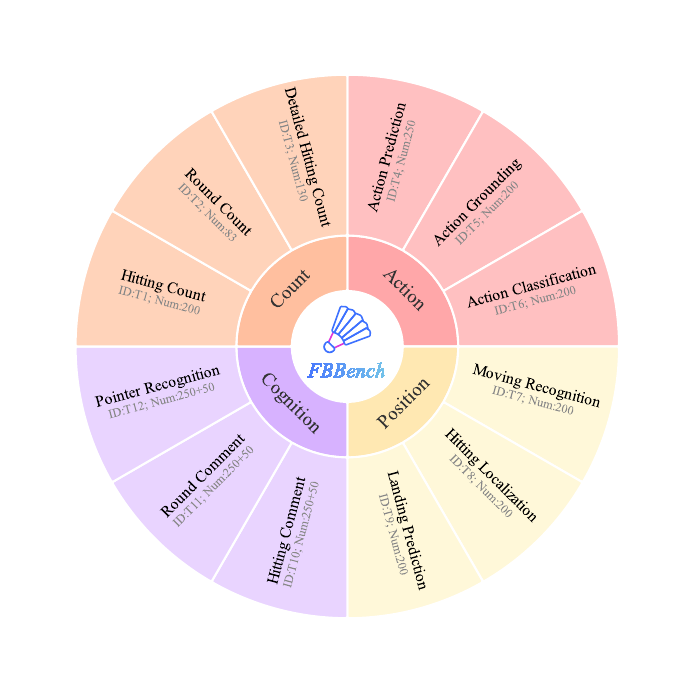}
    \vspace{-1em}
    \caption{Hierarchical task structure of FBBench, showcasing the four primary categories (Count, Action, Position, Cognition) and their respective three sub-categories designed to probe diverse video understanding capabilities.}
    \vspace{-2.5em}
\label{fig:bench_overview}
\end{figure}

\subsection{Dataset Analysis}\label{subsec:Statistics}
The FineBadminton dataset is constructed from 120 singles matches over the past five years. These matches are segmented into 3,215 rally clips, totaling 33,325 strokes. The average duration of each rally is 12.4 seconds. Notably, FineBadminton is, to our knowledge, the first public badminton dataset that features such a detailed multi-level hierarchical annotation scheme combined with rich, free-form textual descriptions for strokes and rallies.

The dataset's hierarchical annotation provides a granular breakdown of badminton gameplay. At the Foundational Actions level, there are 11 primary hit types and 20 subtypes. The shuttlecock's position on the court is classified into 9 distinct regions.

The Tactical Semantics level further enriches the dataset with 3 categories of player actions, 9 strategic classifications, and 6 types of shot characteristics. The outcome of last hit has 3 classes.

In the Decision Evaluation level, individual shots are assigned a quality score from 1 to 7, with hitting comments averaging 37.4 words. The round-level evaluation includes an "Overall Evaluation" averaging 63 words, "Scoring Reason" averaging 41 words, and "Losing Reason" averaging 44 words, providing a comprehensive narrative of the rally's progression and outcome.

\begin{figure*}[t]
    \centering
    \includegraphics[width=\linewidth]{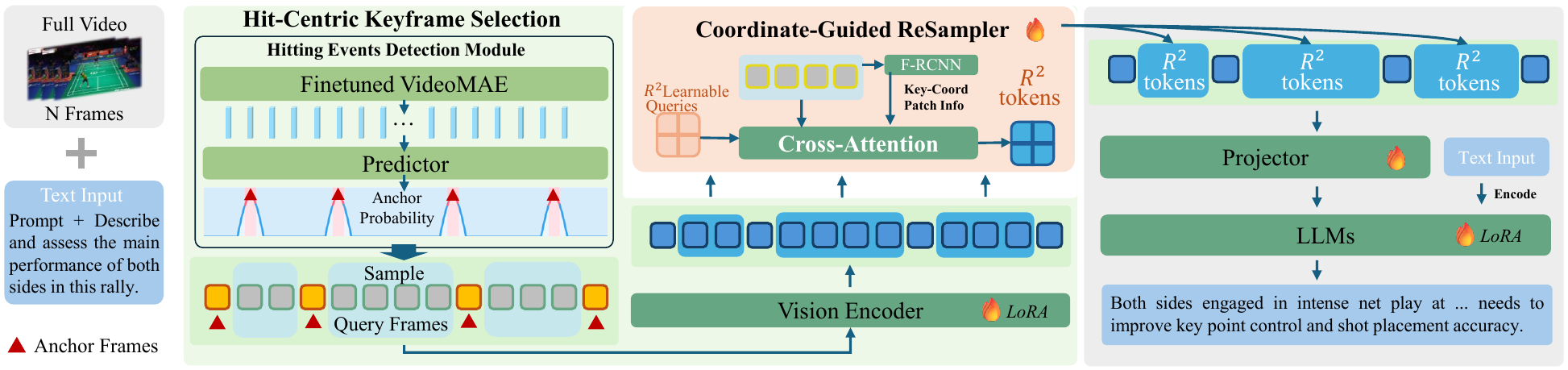}
    \vspace{-2em}
    \caption{
        Architecture of the optimized baseline approach. It employs a two-stage strategy: Hit-Centric Keyframe Selection and Coordinate-Guided Visual Information Condensation, extracting salient information from badminton videos.
    }
\vspace{-1.5em}
\label{fig:agent_architecture}
\end{figure*}

\section{FBBench: A Multi-Faceted Benchmark}

To facilitate a more comprehensive evaluation of badminton video understanding models, we propose \textbf{FBBench}, a novel and challenging benchmark. Built upon the meticulously annotated FineBadminton dataset, FBBench is designed to rigorously assess a model's capabilities in deciphering the intricate dynamics of badminton play, pushing the boundaries beyond simple action recognition.

Specifically, inspired by MVBench~\cite{li2024mvbench}, FBBench emphasizes temporal understanding in fine-grained video analysis, with tasks deliberately crafted to require multi-frame reasoning. FBBench comprises four core domains: \textbf{Count}, \textbf{Action}, \textbf{Position}, and \textbf{Cognition}. Each is further divided into three specialized subcategories, culminating in the 12 distinct task types illustrated in Figure~\ref{fig:bench_overview}. This structure ensures a broad and deep evaluation of model capabilities.

\textbf{Count} tasks assess a model's ability to quantify discrete events, such as strokes per rally or rallies per game, sometimes requiring badminton-specific knowledge like distinguishing shot types.

\textbf{Action} tasks focus on fine-grained motor and tactical recognition, involving  classifying strokes, predicting plausible returns, or temporally localizing key actions within a rally.

\textbf{Position} tasks target spatio-temporal reasoning, including player localization, movement tracking, and shuttlecock landing prediction based on trajectory and context.

\textbf{Cognition} tasks assess higher-level reasoning, challenging models with tasks such as shot quality evaluation, strategic performance inference, and rally outcome determination, all of which require causal inference from extended evidence.

FBBench has 2,563 question-answer pairs. The majority (2,413) are multiple-choice, designed for objective, scalable evaluation, featuring rule-based or LLM-generated plausible distractors and answer length supervision to enhance robustness. Complementing these, 150 open-ended QA tasks assess nuanced descriptive and analytical capabilities. These are evenly distributed across three subcategories of the Cognition task, requiring free-form textual responses for detailed evaluations, scoring or losing reason analysis.

This dual-format strategy, combined with diverse task categories spanning a difficulty gradient—from fundamental perception (e.g., counting, localization) through game mechanics understanding to advanced semantic parsing and strategic reasoning—establishes FBBench as a comprehensive spectrum of challenges. Consequently, FBBench provides a robust platform for evaluating and advancing video understanding models in sports analysis.

\section{Optimized Baseline with Novel Strategies}
\label{sec:baseline}

Given the substantial inter-frame redundancy in badminton videos, which can obscure core actions and reduce MLLM efficiency, a transformation that focuses on salient action segments to create a compact representation is essential. Our optimized baseline approach centers on such an efficient transformation, denoted as $\mathcal{R}$, which converts a raw video $V$ into a compact token sequence $\mathbf{T}$. This representation $\mathbf{T}$ is then fed to an LLM along with a textual query $\mathbf{q}$ to generate a response $\mathbf{r}$: $\mathbf{r} = \text{LLM}(\mathbf{T}, \mathbf{q})$. The transformation $\mathcal{R}$ is specifically designed to address the high temporal density of actions and visual redundancy inherent in badminton videos. 

As illustrated in Figure~\ref{fig:agent_architecture}, $\mathcal{R}$ comprises a carefully designed two-stage process: $\mathcal{R}(V) = \mathcal{C}(\mathcal{S}(V))$. Keyframe Selection ($\mathcal{S}$) firstly pinpoints action-rich pivotal moments to mitigate temporal redundancy. Visual Information Condensation ($\mathcal{C}$) then extracts salient cues from these selected frames, yielding a compact input.

\vspace{-0.5em}
\subsection{Hit-Centric Keyframe Selection}

The Keyframe Selection stage ($\mathcal{S}$) identifies  salient temporal segments.  Its goal is to extract a sparse yet representative set of frames capturing tactically significant moments, critical for comprehensive game understanding.

Our hit-centric strategy $\mathcal{S}$ leverages a hit-event detection module, adapted from the annotation pipeline, utilizing the same VideoMAE model to extract per-frame visual features. These features are subsequently processed through a projection module, which comprises an attention pooling layer and a transformer-based temporal encoder—to output a hit probability $P(\text{hit}|f_i, V)$ for each frame $f_i$ in video segment $V$. Hitting frames are designated as \textbf{anchor frames} ($\mathcal{A}$), representing  significant action points.

These $M$ anchor frames, $\mathcal{A} = \{A_1, \dots, A_M\}$, partition the video into $M-1$ inter-hit segments. To capture crucial pre- and post-hit dynamics, we sample a variable number of \textbf{query frames} from each segment. The total number of selected frames is $M + \sum_{i=1}^{M-1} N_i$, where $N_i$ is the number of query frames in the $i$-th segment.

All selected frames are processed by the MLLM's visual encoder. Each frame is transformed into $K_{enc}$ visual tokens. The complete set of these initial frame token embeddings is denoted as $\mathbf{K}_{emb}$.

\input{tables/quantitative_results}

\vspace{-0.5em}
\subsection{Coordinate-Guided Condensation}

The Visual Information Condensation stage ($\mathcal{C}$) distills $\mathbf{K}_{emb}$ into the compact token sequence $\mathbf{T} = \mathcal{C}(\mathbf{K}_{emb})$, to minimize static background redundancy while preserving crucial semantic cues.

Anchor frame tokens $\mathbf{E}_{A_j}$ are preserved. For query frame tokens $\mathbf{X}_{i:i+1}$, we employ a ReSampler module, inspired by the Perceiver architecture. This module uses $R$ learnable queries $\mathbf{Q}_{\text{learn}} \in \mathbb{R}^{R \times D}$ to distill information via cross-attention.

This resampling is \textbf{coordinate-guided}. We detect salient game elements (ball, players, court keypoints) in query frames using a FastRCNN-based object detector. These spatial detections guide the ReSampler. Let $\mathbf{q}'$ be processed learnable queries, and $\mathbf{k}'^{(i)}, \mathbf{v}'^{(i)}$ be key and value sequences for the $i$-th segment, derived from $\mathbf{Q}_{\text{learn}}$ and $\mathbf{X}_{i:i+1}$ respectively. The attention scores for segment $i$ are:
$$\text{Score}^{(i)} = \frac{\mathbf{q}' (\mathbf{k}'^{(i)})^T}{\sqrt{D}} + \mathbf{B}_{\text{coord}}^{(i)}$$
where $D$ is the dimension of key vectors. The coordinate-bias tensor $\mathbf{B}_{\text{coord}}^{(i)} \in \mathbb{R}^{B \times R \times (N_i \cdot K_{enc})}$ has element $(b, p, q)$ set to a scalar hyperparameter $\alpha > 0$ if the $q$-th input token in $\mathbf{X}_{i:i+1}$ spatially aligns with a detected key game element, and $0$ otherwise. This additive bias effectively steers the ReSampler's attention towards these critical visual regions.The condensed representation for the query segment is $\mathbf{R}_{i:i+1} = \mathrm{Softmax}(\text{Score}^{(i)}) \mathbf{v}'^{(i)} \in \mathbb{R}^{B \times R \times D}$.

The final token sequence $\mathbf{T}$ concatenates anchor frame tokens and condensed query segment representations temporally:
$$\mathbf{T} = \text{Concat} \left( \mathbf{E}_{A_1}, \mathbf{R}_{1:2}, \mathbf{E}_{A_2}, \dots, \mathbf{R}_{M-1:M}, \mathbf{E}_{A_M} \right)$$
This results in $\mathbf{T} \in \mathbb{R}^{B \times (M \cdot K_{enc} + (M-1) \cdot R) \times D}$, emphasizing pivotal hits and salient inter-hit cues for LLM processing.

\noindent \textbf{Fine-Tuning.} The training set consists of approximately 60,000 QA pairs. We construct these pairs from the FineBadminton, designing question types based on the categories outlined in the FBBench. This results in 10\% multiple-choice questions and the remainder open-ended questions, all of which are formulated and trained as standard sequence-to-sequence generation tasks for the LLM. Within the MLLM, the visual encoder and the LLM were fine-tuned using LoRA, whereas the ReSampler and the hit-centric projection module underwent full-parameter training.

\vspace{-0.5em}
\section{Experiments}
\subsection{Experimental Settings}
\textbf{Models and Methodologies.} We evaluated several state-of-the-art models on our FBBench. These include prominent  commercial models such as Gemini 2.5 Pro\cite{team2023gemini}, Doubao 1.5 Pro\cite{doubao}, and GPT-4.1\cite{gpt4.1}, as well as leading open-source models such as Qwen2.5VL-7B\cite{bai2025qwen2}, and VideoLLaMA3-7B\cite{zhang2025videollama}. All models were benchmarked using two distinct video frame processing approaches: (1) a baseline uniform sampling at 2 FPS, and (2) our proposed $\mathcal{S}$ strategy.

\textbf{Experimental Details.} 
For the commercial models, accessed via APIs, we evaluated the impact of our $\mathcal{S}$ strategy.
For the open-source models, we applied our optimized baseline incorporating both the $\mathcal{S}$ and $\mathcal{C}$ strategies, and trained these models on 4 A100-40G GPUs. To match the frame count of the uniform baseline, uniformly sampled inter-keyframe frames were added when $\mathcal{S}$ yielded fewer frames.

\textbf{Metrics.} 
Performance on multiple-choice questions was measured by accuracy, calculated based on the total number of correct answers. To mitigate potential biases from option ordering, we implemented a re-evaluation loop where options were shuffled for each assessment.
For open-ended questions, we utilized GPT-4.1 as an evaluator. Each response was scored by GPT-4.1 on a scale of 0 to 10 points, based on relevance, correctness, and completeness, against a reference answer. The overall performance for open-ended tasks was reported as the cumulative score across all such questions.

\vspace{-0.5em}
\subsection{Quantitative Evaluation}
Quantitative results on FBBench are presented in Table~\ref{tab:quantitativeeva}. Among commercial models, performance rankings aligned with general expectations: Gemini 2.5 Pro led, followed by Doubao 1.5 Pro, and then GPT-4.1. The benchmark's difficulty is evident, as even the top-performing Gemini 2.5 Pro (with $\mathcal{S}$) achieved an overall multiple-choice score of 932 (representing 38.6\% accuracy), indicating significant headroom for improvement. Across all models, tasks requiring deep domain-specific knowledge (e.g., T3, T6), subtle action discrimination (e.g., T1, T7), or complex reasoning (e.g., T10-T12) proved particularly challenging. Notably, our strategy $\mathcal{S}$ consistently enhanced the performance of these commercial models over uniform sampling with an equivalent frame count.

Remarkably, the smaller 7B open-source models, which initially performed below 25\% overall accuracy, demonstrated substantial improvements after fine-tuning with $\mathcal{S}$ and $\mathcal{C}$ strategies. The optimized Qwen2.5VL-7B achieved an overall 42.06\% multiple-choice accuracy, surpassing all evaluated commercial models including the leading Gemini 2.5 Pro (with $\mathcal{S}$). Furthermore, their performance on open-ended QA tasks (e.g., Qwen2.5VL-7B: 655) also became highly competitive, approaching the levels of top commercial models.

\input{tables/ablation_studies}

\subsection{Ablation Studies}

To isolate and validate the effectiveness of our proposed $\mathcal{S}$ and $\mathcal{C}$ strategies, we conducted ablation studies on VideoLLaMA3. The results are presented in Table~\ref{tab:ablation_studies}.

Fine-tuning VideoLLaMA3 on our FineBadminton subset substantially boosted its overall score from 359 to 627, highlighting the critical role of domain-specific fine-tuning. Further enhancements were achieved by integrating our proposed strategies: the $\mathcal{S}$ strategy (with $FT$) increased the score to 737, and the $\mathcal{C}$ strategy (with $FT$) reached 856. Notably, our full model, combining all components, achieved the highest score of 954, demonstrating the synergistic benefits of domain-specific fine-tuning with our tailored video processing for effective fine-grained analysis.

\vspace{-0.5em}
\section{Conclusion}
In this paper, we introduced \textbf{FineBadminton}, a novel, large-scale dataset for fine-grained badminton video understanding, distinguished by its comprehensive multi-level semantic annotation hierarchy. The dataset's construction is supported by an MLLM-driven annotation pipeline with expert refinement. We also presented \textbf{FBBench}, a challenging benchmark for rigorous assessment of nuanced video understanding. While our video processing strategies enhance MLLM performance, evaluations on FBBench demonstrate current models struggle with the complex dynamics and tactics of high-speed sports videos. We believe this work provides a crucial foundation to spur advancements in fine-grained video analysis.

\begin{acks}
This work was supported in part by the National Natural Science Foundation of China under Grant 62376069, in part by the Young Elite Scientists Sponsorship Program by CAST under Grant 2023QNRC001, in part by Guangdong Basic and Applied Basic Research Foundation under Grant 2024A1515012027, in part by the Shenzhen Science and Technology Program under Grant KQTD2024\\0729102207002 and Grant ZDSYS20230626091203008, and in part by Jiangsu Science and Technology Major Program under Grant BG2024041.
\end{acks}

\bibliographystyle{ACM-Reference-Format}
\bibliography{ref}

\appendix


\section{The FineBadminton}
We introduce FineBadminton, a comprehensive, and multi-modal dataset designed for fine-grained analysis of badminton matches. As shown in the comparative analysis in Table~\ref{tab:dataset_comparison}, while numerous datasets for sports video analysis exist, there is a notable gap, particularly in the realm of badminton. Previous badminton datasets like ShuttleNet, ShuttleSet, and ShuttleSet22 have primarily focused on stroke detection and rally localization, lacking the detailed, multi-faceted annotations necessary for deeper semantic understanding.

\begin{table*}[b!]
\centering
\caption{Comparison of sports video analysis datasets. Our proposed dataset, FineBadminton, is highlighted in blue.
Abbreviations: HA (Hierarchical Annotation), TA (Textual Annotation), UED (Uses Existing Data, e.g., scores, commentary), AQA (Action Quality Assessment), FGL (Fine-grained Labels).}
\label{tab:dataset_comparison}
\vspace{-6pt}
    \setlength{\tabcolsep}{0.15cm} 
    \renewcommand{\arraystretch}{1.1} 
    \scalebox{1}{
    \begin{tabular}{c c ccc | ccccc}
    \toprule
    \textbf{Dataset} & \textbf{Sport} & \textbf{Videos} & \textbf{Rallies/Clips} & \textbf{Strokes/Actions/Evts.} & \textbf{HA} & \textbf{TA} & \textbf{UED} & \textbf{AQA} & \textbf{FGL} \\
    \midrule
    Finegym(CVPR'20) & Gymnastics & 303 & 4,883 & 32,697 & \ding{52} & \ding{56} & \ding{52} & \ding{52} & \ding{52} \\
    SoccerNetV2(CVPR'21) & Soccer & 500 & - & 110,458 & \ding{52} & \ding{56} & \ding{56} & \ding{56} & \ding{52} \\
    FineDiving(CVPR'22) & Diving & 135 & - & 3,000 & \ding{56} & \ding{56} & \ding{56} & \ding{52} & \ding{52} \\
    ShuttleNet(AAAI'22) & Badminton & 75 & 4,325 & 43,191 & \ding{56} & \ding{56} & \ding{56} & \ding{56} & \ding{56} \\
    ShuttleSet(KDD'23) & Badminton & 44 & 3,685 & 36,492 & \ding{56} & \ding{56} & \ding{56} & \ding{56} & \ding{56} \\
    ShuttleSet22(IJCAI'23) & Badminton & 58 & 3,992 & 33,612 & \ding{56} & \ding{56} & \ding{56} & \ding{56} & \ding{56} \\
    Logo(CVPR'23) & Artistic Swim. & 200 & - & 15,764 & \ding{52} & \ding{56} & \ding{52} & \ding{52} & \ding{52} \\
    FineSports(CVPR'24) & Basketball & - & 10,000 & 16,000 & \ding{52} & \ding{56} & \ding{56} & \ding{56} & \ding{52} \\
    P2anet(TOMM'24) & Table Tennis & 200 & 2,721 & 139,075 & \ding{56} & \ding{56} & \ding{56} & \ding{56} & \ding{56} \\
    SportsHHI(CVPR'24) & B-ball/V-ball & 160 & - & 50,649 & \ding{52} & \ding{56} & \ding{56} & \ding{56} & \ding{52} \\
    F$^3$Set(ICLR'25) & various & 114 & 11,584 & 42,846 & \ding{52} & \ding{56} & \ding{56} & \ding{56} & \ding{52} \\
    \rowcolor{blue!15}
    \textbf{FineBadminton} & Badminton & 120 & 3,215 & 33,325 & \ding{52} & \ding{52} & \ding{56} & \ding{52} & \ding{52} \\
    \bottomrule
    \end{tabular}
    }
\end{table*}

\vspace{-1em}

\section{The FBBench}

\subsection{Evaluation Protocol and Prompting}

The evaluation protocol for the FBBench is tailored to its two distinct question formats. For the multiple-choice tasks, models are instructed to respond with a single letter corresponding to their chosen option. A structured prompt is employed: "This is a segment from a badminton match. `top' refers to the player at the top of the screen, and `bottom' refers to the player at the bottom. \{Question\} Watch the video carefully, paying attention to the rally's progression, and the players' actions and postures. Based on your observation, select the option that most accurately answers the question. Respond with a single option letter."  Robust answer extraction is ensured through predefined templates, supplemented by GLM-4-air\cite{glm2024chatglm} for non-standard responses. To enhance rigor and mitigate guessing, a re-evaluation loop with shuffled options is implemented for initially correct answers.


\vspace{-0.5em}

\subsection{Questions Mapping}
\input{tables/question_map}

\end{document}

%% file: tables/quantitative_results.tex
\begin{table*}[t]
\centering
\caption{
Quantitative Comparisons on FBBench. Numbers indicate correctly answered questions. For open-ended questions (marked with *), the number represents total score (0–10 per response). $\mathcal{S}$ and $\mathcal{C}$ denote Hit-Centric Keyframe Selection and Coordinate-Guided Condensation in Section \ref{sec:baseline}. T$i$ denotes the $i$-th type of question, refer to Figure~\ref{fig:bench_overview} for the detailed mapping.
}
\renewcommand{\arraystretch}{0.9}
\vspace{-6pt}
\label{tab:quantitativeeva} 
    \setlength{\tabcolsep}{0.10cm} 
    \renewcommand{\arraystretch}{1.0} 
    \resizebox{.98\textwidth}{!}{ 
    \begin{tabular}{c|*{3}{c}*{3}{c}*{3}{c}*{6}{c}|*{2}{c}}
    \toprule[1pt]
    \multirow{2}{*}{\textbf{Model}} 
    & \multicolumn{3}{c|}{\textbf{Count}} 
    & \multicolumn{3}{c|}{\textbf{Action}} 
    & \multicolumn{3}{c|}{\textbf{Position}} 
    & \multicolumn{6}{c|}{\textbf{Cognition}} 
    & \multicolumn{2}{c}{\textbf{Overall}} \\
    \cmidrule(lr){2-4} \cmidrule(lr){5-7} \cmidrule(lr){8-10} \cmidrule(lr){11-16} \cmidrule(lr){17-18}
    & \textbf{T1} & \textbf{T2} & \textbf{T3} 
    & \textbf{T4} & \textbf{T5} & \textbf{T6} 
    & \textbf{T7} & \textbf{T8} & \textbf{T9} 
    & \textbf{T10} & \textbf{T10*} & \textbf{T11} & \textbf{T11*} & \textbf{T12} & \textbf{T12*}
    & \textbf{Choice} & \textbf{Open-ended} \\
    \midrule
    \textbf{Full Score} & 200 & 83 & 130 & 250 & 200 & 200 & 200 & 200 & 200 & 250 & 500 & 250 & 500 & 250 & 500 & 2413 & 1500 \\
    \midrule
    \multicolumn{18}{c}{\textbf{Commerical Models}} \\
    \midrule
    Gemini 2.5 Pro & 69 & \textbf{33} & 34 & 67 & 70 & 44 & 54 & 63 & 45 & 124 & 192 & 114 & 214 & 155 & 325 & 872 (36.14\%) & 731 (48.73\%) \\
    Gemini 2.5 Pro + $\mathcal{S}$ & \textbf{79} & 28 & 35 & \textbf{68} & \textbf{71} & 54 & \textbf{58} & \textbf{70} & \textbf{53} & 131 & 202 & 117 & 222 & \textbf{168} & \textbf{335} & \textbf{932 (38.62\%)} & \textbf{759 (50.60\%)} \\
    Doubao 1.5 pro & 71 & 23 & 45 & 55 & 50 & 36 & 40 & 61 & 52 & \textbf{133} & 165 & 112 & 213 & 133 & 286 & 811 (33.61\%) & 664 (44.27\%) \\
    Doubao 1.5 pro + $\mathcal{S}$ & 71 & 30 & \textbf{48} & 56 & 57 & 39 & \textbf{48} & \textbf{68} & 45 & 130 & 188 & \textbf{121} & \textbf{227} & 143 & 294 & 856 (35.47\%) & 709 (47.27\%) \\
    GPT-4.1 & 47 & 12 & 40 & \textbf{76} & 55 & 60 & 40 & 30 & 48 & 110 & 183 & 98 & 217 & 128 & 310 & 743 (30.79\%) & 710 (47.33\%) \\
    GPT-4.1 + $\mathcal{S}$ & 46 & 17 & 41 & 70 & 61 & \textbf{64} & 44 & 39 & 49 & 112 & 186 & 100 & 213 & 138 & 316 & 782 (32.41\%) & 715 (47.67\%) \\
    \midrule
    \multicolumn{18}{c}{\textbf{Open-Source Models}} \\
    \midrule
    
    Qwen2.5VL-7B & 48 & 19 & 30 & 50 & 39 & 33 & 52 & 41 & 41 & 60 & 110 & 62 & 79 & 82 & 46 & 557 (23.08\%) & 235 (15.67\%) \\
    Qwen2.5VL-7B + $\mathcal{S}$ + $\mathcal{C}$ & \textbf{70} & \textbf{24} & \textbf{36} & \textbf{92} & 44 & \textbf{87} & \textbf{56} & 55 & \textbf{98} & \textbf{170} & \textbf{172} & \textbf{142} & \textbf{205} & \textbf{141} & \textbf{278} & \textbf{1015 (42.06\%)} & \textbf{655 (43.67\%)} \\
    VideoLLaMA3-7B & 42 & 10 & 20 & 17 & 44 & 24 & 30 & 20 & 12 & 45 & 95 & 55 & 66 & 40 & 47 & 359 (14.88\%) & 208 (13.87\%) \\
    VideoLLaMA3-7B + $\mathcal{S}$ + $\mathcal{C}$ & 52 & \textbf{24} & 25 & 85 & \textbf{54} & 80 & 40 & \textbf{68} & 96 & 165 & 167 & 130 & 198 & 135 & 264 & 954 (39.54\%) & 629 (41.93\%) \\
\bottomrule[1pt]
\end{tabular}
}
\vspace{-6pt}
\end{table*}

%% file: tables/ablation_studies.tex
\begin{table}[t]
\centering
\caption{Ablations on FBBench Multi-Choice QA. $FT$ denotes finetuning. The \setlength{\fboxsep}{1pt}\colorbox{blue!20}{blue background} indicates our approach.}
\label{tab:ablation_studies}
\vspace{-6pt}

    \setlength{\tabcolsep}{0.10cm} 
    \renewcommand{\arraystretch}{1.0} 

    \scalebox{0.88}{
    \begin{tabular}{*{3}{c}|*{4}{c}c} 
    \toprule
     $\mathcal{S}$ & $\mathcal{C}$ & $FT$ & \textbf{Count} & \textbf{Action} & \textbf{Localization} & \textbf{Cognition} & \textbf{Overall} \\
    \midrule
     \ding{56} & \ding{56} & \ding{56} &  72 &  85 &  62 &  140 & 359 \\
     \ding{56} & \ding{56} & \ding{52} &  70 &  132 &  124 &  301 & 627 \\
     \ding{52} & \ding{56} & \ding{52} &  75 &  153 &  147 &  362 & 737 \\
     \ding{56} & \ding{52} & \ding{52} &  81 &  195 &  186 &  394 & 856 \\
    \rowcolor{blue!15}
     \ding{52} & \ding{52} & \ding{52} &  \textbf{101} &  \textbf{219} & \textbf{204} & \textbf{430} & \textbf{954} \\
    \bottomrule
    \end{tabular}
    } 
    \vspace{-0.7em}
\end{table}

%% file: tables/question_map.tex
\begin{table}[H]
\centering
\caption{Mapping of Spatial and Temporal Dimensions to Task IDs. This table outlines the different categories and their corresponding identifiers used in our analysis.}
\label{tab:question_map_styled} 
\vspace{-6pt} 

\setlength{\tabcolsep}{4pt}

\renewcommand{\arraystretch}{1.0}

\scalebox{0.90}{
\begin{tabular}{lll}
\toprule
\textbf{Spatial Dimension} & \textbf{Temporal Dimension} & \textbf{ID} \\
\midrule
\multirow{3}{*}{Count} & Hitting Count & T1 \\
                       & Round Count & T2 \\
                       & Detailed Hitting Count & T3 \\
\midrule
\multirow{3}{*}{Action} & Action Prediction & T4 \\
                       & Action Grounding & T5 \\
                       & Action Classification & T6 \\
\midrule
\multirow{3}{*}{Position} & Moving Recognition & T7 \\
                        & Hitting Localization & T8 \\
                        & Landing Prediction & T9 \\
\midrule
\multirow{3}{*}{Cognition} & Hitting Comment & T10 \\
                         & Round Comment & T11 \\
                         & Pointer Recognition & T12 \\
\bottomrule
\end{tabular}
} 
\end{table}